\documentclass{emulateapj}

\usepackage{graphicx}

\shorttitle{Chandrasekhar-Fermi Method}
\shortauthors{Cho \& Yoo}

\begin{document}

\title{ A technique for constraining the driving scale of turbulence and a modified
 Chandrasekhar-Fermi method }
\author{ Jungyeon Cho, Hyunju Yoo}
\affil{Department of Astronomy and Space Science, Chungnam National University, Daejeon, Korea; jcho@cnu.ac.kr, hyunju527@gmail.com}

\begin{abstract}
The Chandrasekhar-Fermi method is a powerful technique for estimating
the strength of the mean magnetic field projected on the plane of the sky.
In this paper, 
we present a technique for improving the
 Chandrasekhar-Fermi method, in which we take into
account the averaging effect arising from independent eddies along the line of sight .
In the conventional Chandrasekhar-Fermi method,
the strength of fluctuating magnetic field divided by $\sqrt{4 \pi \bar{\rho}}$, where
$\bar{\rho}$ is average density, is assumed to be comparable to the line-of-sight velocity dispersion.
 This however is not true when the driving scale of turbulence $L_f$,
 i.e.~the outer scale of turbulence,  is
 smaller than the size of the system along the line of sight $L_{los}$.
 In fact, the conventional Chandrasekhar-Fermi method over-estimates the 
 strength of the mean plane-of-the-sky magnetic field
 by a factor of $\sim \sqrt{ L_{los}/L_f}$.
 We show that the standard deviation of  centroid velocities divided by the average 
line-of-sight velocity dispersion is a good measure of $\sqrt{ L_{los}/L_f}$, which
enables us to propose a modified Chandrasekhar-Fermi method.

\end{abstract}
\maketitle

\section{Introduction}
Magnetic fields play important roles in many astrophysical environments.
However, measuring the strength of them is a very challenging problem.
The Chandrasekhar-Fermi method (Chandrasekhar \& Fermi 1953; 
hereinafter the CF method) 
is a simple and powerful technique that
can measure the strength of regular magnetic field perpendicular to the
line of sight, i.e.~the component of the mean magnetic field projected on the plane of the sky.

The idea of the CF method is simple. 
Let us consider a fluid filled with Alfv\'en waves or Alfv\'enic turbulence.
In Alfv\'enic disturbances, the r.m.s. fluctuation of magnetic field ($\delta b$) and
the r.m.s. velocity ($\delta  v$) are related by
\begin{equation}
    \frac{\delta b}{\sqrt{4 \pi \bar{\rho}} } \sim \delta v \mbox{~~~or~~} 
   1 \sim  \sqrt{ 4 \pi \bar{\rho} } \frac{ \delta v }{ \delta b },
\end{equation}
where $\bar{\rho}$ is average density.
If we multiply both sides by the mean plane-of-the-sky magnetic field $B_{0,sky}$,
we obtain
\begin{equation}
  B_{0,sky} \sim \sqrt{ 4 \pi \bar{\rho} } \frac{ \delta v }{ \delta b/B_{0,sky} }. \label{eq:2}
\end{equation}
If the velocity fluctuation ($\delta v$) and the magnetic field fluctuation ($\delta b$) 
are isotropic,
then we may write $\delta v_{los}/\delta b_{\bot, sky} \sim \delta v/\delta b$, where 
$\delta v_{los}$ is the line-of-sight velocity dispersion and 
$\delta b_{\bot, sky}$ is the r.m.s. fluctuation of the
plane-of-the-sky magnetic field that is perpendicular to 
$B_{0, sky}$. Therefore, the equation above becomes
\begin{equation}
   B_{0,sky} 
   =\xi \sqrt{ 4 \pi \bar{\rho} } \frac{ \delta v_{los} }{ \delta b_{\bot, sky}/B_{0,sky} } 
     \sim \xi \sqrt{ 4 \pi \bar{\rho} } \frac{ \delta v_{los} }{\delta \phi },   \label{eq:trad}
\end{equation}
where $\delta \phi$ is the variation of the angle between the plane-of-the-sky magnetic field 
and the mean plane-of-the-sky magnetic field $\bold{B}_{0,sky}$, and
we use 
\begin{equation}
    \delta \phi \sim \tan (\delta \phi)= \delta b_{\bot, sky}/B_{0,sky}.  \label{eq:phi}
\end{equation}
The factor $\xi$ is a correction factor, which is usually taken as $\sim$0.5 
(Ostriker et al 2001; Padoan et al. 2001; Heitsch et al. 2001).
We can obtain  $\delta \phi$ from observations of star-light polarization or polarized far-infrared emission from magnetically aligned 
dust grains and we can measure $\delta v_{los}$ from
the width of an optically thin molecular emission line (see, for example, 
Gonatas et al. 1990; Lai et al. 2001; Di Francesco et al. 2001;
Crutcher et al. 2004; Girart et al. 2006; Curran \& Chrysostomou 2007; 
Heyer et al 2008; Mao et al. 2008; Tang et al. 2009; Sugitani et al. 2011).
Further elaboration of the CF method has been made by many researchers 
(Zweibel 1990; Myers \& Goodman 1991; Zweibel 1996; Ostriker et al. 2001; 
Heitsch et al. 2001; Padoan et al. 2001; Kudoh \& Basu 2003; Wiebe \& Watson 2004;
Falceta-Gon{\c c}alves et al. 2008; Hildebrand et al. 2009; Houde et al 2009).

Reduction of $\delta \phi$ due to averaging effects is of great importance
in our paper.
Roughly speaking, two main averaging effects exist.
First, averaging along the line of sight can reduce $\delta \phi$.
That is, if there are more than one independent turbulent eddies along the line of sight,
the measured value of $\delta \phi$ will be reduced (see Myers \& Goodman 1991;
Zweibel 1996; Houde et al 2009).
Second, averaging the polarization angle
within the telescope beam can also reduce $\delta \phi$
(see Heitsch et al. 2001; Wiebe \& Watson 2004; Falceta-Gon{\c c}alves et al. 2008;
Houde et al. 2009).
The averaging effects 
in general make the step from Equation (\ref{eq:2}) to Equation (\ref{eq:trad})
inaccurate.
In the presence of the averaging effects, the CF method tends to overestimate
$B_{0,sky} $.

In this paper, we focus on the averaging effect along the line of sight and
propose a simple method to compensate the effect.
Using 3-dimensional direct magnetohydrodynamic (MHD) turbulence simulations, 
we test the proposed technique.
In \S2, we describe our numerical method and explain  in detail how averaging along the
line of sight makes the conventional CF method overestimate $B_{0,sky} $.
At the end of \S2, we describe our new technique.
In \S3, we present results of our numerical simulations.
In \S4, we give discussions and summary.

\section{Numerical Method and Theoretical Considerations}
\subsection{Numerical code}
\label{sect:code}
To obtain turbulence data, we solve the following compressible MHD equations 
in a periodic box of size $2\pi$
using
an Essentially Non-Oscillatory scheme (see Cho $\&$ Lazarian 2002):
\begin{eqnarray}
{\partial \rho    }/{\partial t} + \nabla \cdot (\rho {\bf v}) =0,  \\
{\partial {\bf v} }/{\partial t} + {\bf v}\cdot \nabla {\bf v} 
   +  \rho^{-1}  \nabla(C_s^2\rho)\nonumber\\
   - (\nabla \times {\bf B})\times {\bf B}/4\pi \rho ={\bf f},  \\
{\partial {\bf B}}/{\partial t} -
     \nabla \times ({\bf v} \times{\bf B}) =0, 
\end{eqnarray}
with $\nabla$$\cdot$$\bf B$$=$0 and an isothermal equation of state $P$=$C^{2}_{s}\rho$, 
where  $C_{s}$ is the sound speed and $\rho$ is density. 
Here $\bf{v}$ is the velocity, $\bf{B}$ is the magnetic field, 
and $\bf{f}$ is the driving force. We use $512^3$ grid points.
In our simulations, $C_s=0.1$, $\bar{\rho}=1$, and $B_0/\sqrt{ 4 \pi \bar{\rho}}=1$.
In all simulations, the r.m.s. velocity $v_{rms}$ is between $\sim$0.7 and $\sim$0.8, and
the sonic Mach number is 
$M_s \equiv v_{rms}/C_s\sim 7$.
Since the Alfv\'en speed of the mean field ($V_A=B_0/\sqrt{ 4 \pi \bar{\rho}}$) is 1,
the Alfv\'en Mach number is $M_A\equiv v_{rms}/V_A \sim 0.7$, which
means that turbulence considered in this paper is sub-Alfv\'enic.

\subsection{Forcing}
In this work, we drive turbulence in Fourier space and consider only solenoidal ($\nabla \cdot {\bf f}=0$) forcing. 
We use  
$\sim$100  forcing components isotropically distributed in the range $k_f/1.3\lesssim$k$\lesssim 1.3k_f$, where
$k$ is the wavenumber and $k_f$ varies from simulation to simulation (see Table 1).
Therefore, the peak of energy injection occurs at k $\sim k_f$.
    More detailed descriptions on forcing can be found in Yoo \& Cho (2014).

\begin{figure}
\center
\includegraphics[width=0.48\textwidth]{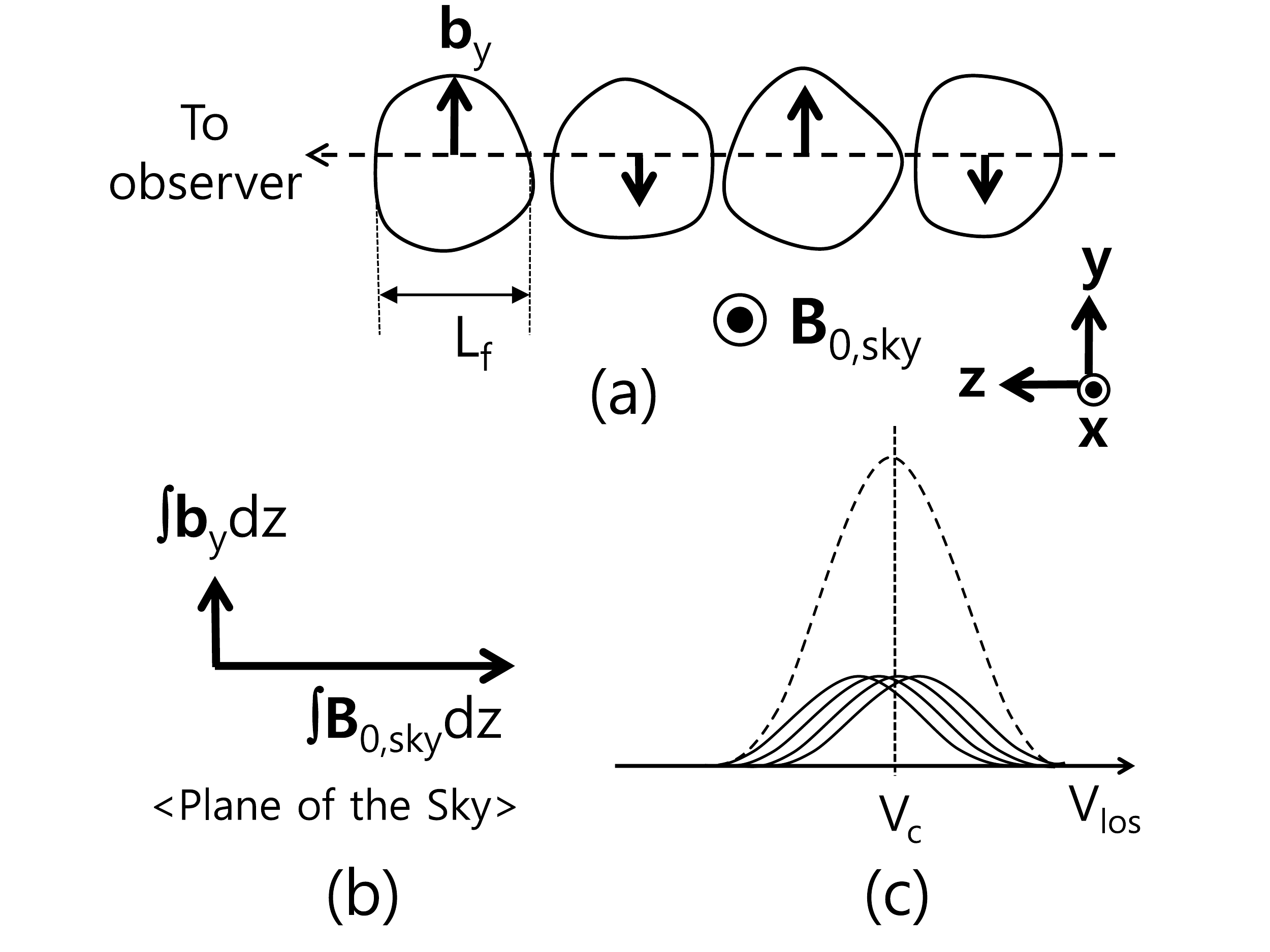}  
\caption{Effects of averaging along the line of sight.
   (a) If the driving scale of turbulence is $L_f$ in a strongly magnetized medium, 
        we have $\sim L_{los}/L_f ~(\equiv N)$
        independent eddies
         along the line of sight. Here $L_{los}$ is the size of the system along the 
         line of sight. 
         The field ${\textbf B}_{0,sky}$ denotes the mean plane-of-the-sky magnetic field.
         The $y$-component of magnetic field in each eddy is likely to be random.
   (b) The $x$-component of observed magnetic field is equal to
        $\int B_{0,sky} dz \sim B_{0,sky} L_{los}$, 
        if we ignore contribution from random magnetic field along $x$-direction.
       The $y$-component of observed magnetic field is
        $\sim \int b_y dz \sim (b_y L_f)\sqrt{L_{los}/L_f}$.
   (c) The solid lines denote line profiles from individual eddies and the dashed line
       stands for observed line profile. The mean velocity of each eddy
       can be different from each other. The observed  centroid velocity $V_c$ is
       approximately the average of the mean velocities of individual  eddies along the line of sight.
 }
\label{fig:1}
\end{figure}

\subsection{Theoretical Considerations: The Effects of Averaging along a Line of Sight}
Our main concern is to investigate the effect of the driving scale on the CF method.
For this purpose, we change the driving scale by changing the driving wavenumber $k_f$.

If the driving scale is $L_f$, then there are $\sim L_{los}/L_f ~(\equiv N)$ large-scale eddies 
(i.e.~largest energy-containing eddies) along a line of sight, where $L_{los}$
is the size of the system along the line of sight (see Figure \ref{fig:1}(a))\footnote{
     In our simulations, $L_{los}=2 \pi$ and $L_f \sim 2\pi /k_f$.
    Therefore, if the driving wavenumber is $k_f$, then
    the number of independent eddies along a line of sight is 
    $N=L_{los}/L_f \sim k_f$.}.
In this case, 
what will be the strength of observed magnetic field projected
on the plane of the sky?
If we take a coordinate system as shown in Figure \ref{fig:1}(a), the plane of the
sky is parallel to the $xy$-plane.
Let us consider the $x$ and $y$-components of magnetic field separately. 
Here $x$ and $y$-directions are
parallel and perpendicular to the mean plane-of-the-sky magnetic field, respectively.
We can write
\begin{eqnarray}
   B_{x, obs}\propto \int_0^{L_{los}} B_x dz \sim B_{0,sky} L_{los}, \\
   b_{y, obs}\propto \int_0^{L_{los}} b_y dz \sim b_y L_f \sqrt{\frac{L_{los}}{L_f}}
                 = b_y    \sqrt{L_f L_{los}},   \label{eq:by}
\end{eqnarray}
where we ignore the contribution of random magnetic field 
in the integration of $B_x$ ($\equiv B_{0,sky}+b_x$) and
assume that each large-scale eddy contributes randomly 
in the integration of $b_y$ (see Figure \ref{fig:1}(b))\footnote{
   The length-scale $L_f$ in Equation (\ref{eq:by}) should be
   the coherence scale of magnetic field (see Cho \& Ryu 2009).
   In this paper, we assume that the coherence scale coincides with the
   driving scale of turbulence, which is a very good approximation for trans-Alfv\'enic or
   sub-Alfv\'enic turbulence, i.e.~turbulence in which the r.m.s. velocity is similar to or less
   than the Alfv\'en speed of the mean magnetic field.
   In sub-Alfv\'enic turbulence, the magnetic 
   energy spectrum peaks at the driving scale and, therefore, the coherence scale of
   magnetic field should be very close to the driving scale of turbulence.}.
Therefore, the variation of the 
angle between the magnetic field projected on the plane of the sky and 
the mean plane-of-the-sky magnetic field is given by
\begin{equation}
  \delta \phi \sim \tan (\delta \phi) \sim \frac{ b_{y,obs} }{ B_{x,obs} } \sim 
  \frac{ b_y }{ B_0 } \sqrt{ L_f/L_{los} },
\end{equation}
which is $( L_{los}/L_f )^{1/2}$ times smaller than the conventional estimate of
$\delta \phi$ (see Equation (\ref{eq:phi})).

\begin{figure*}
\center
\includegraphics[width=0.80\textwidth]{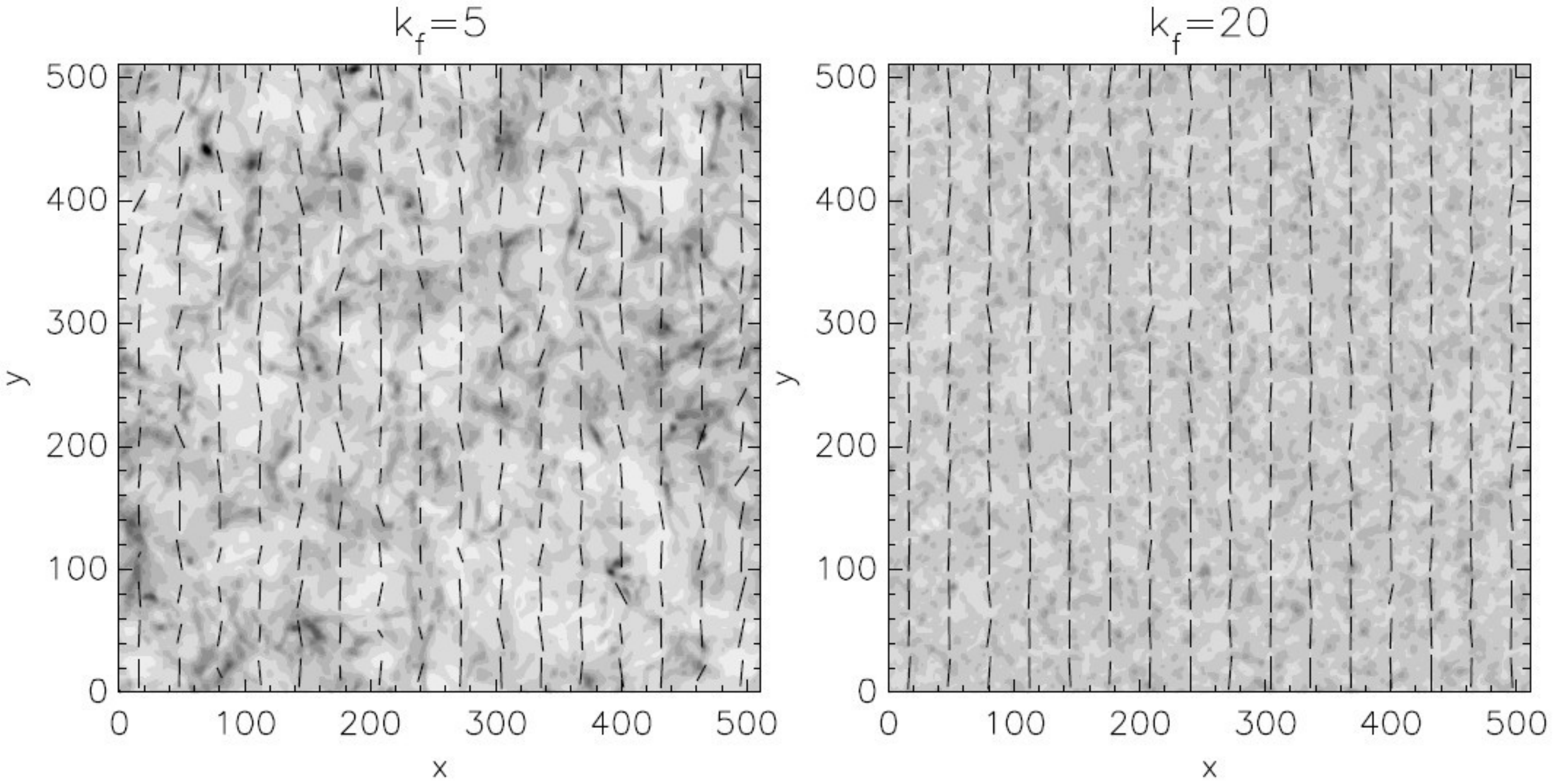}  
\caption{Reduction of variation in polarization angle $\delta \phi$ due to
  averaging effect along the line of sight.
The driving wavenumbers, hence the number of independent eddies
along a line of sight, are different in the left and right panels.
 {\it Left:} The driving wavenumber is 
       $\sim5$ 
     and therefore there are approximately 
     5
     independent eddies along a line of sight.
 {\it Right:} The driving wavenumber is $\sim20$ and 
   there are approximately 20 independent eddies along a line of sight.
   Variation in polarization angle is smaller in the right panel due to larger number
   of independent eddies along the line of sight.
The contours represent intensity of FIR/sub-mm emission from magnetically aligned dust.
 The polarization maps are taken after saturation of turbulence. The left and the right 
  panels are from Runs 
  KF5
  and KF20, respectively.
  The projections are done along a line of sight perpendicular to
  the mean field ${\textbf B}_0$.}
\label{fig:2}
\end{figure*}

\begin{figure*}
\includegraphics[width=0.5\textwidth]{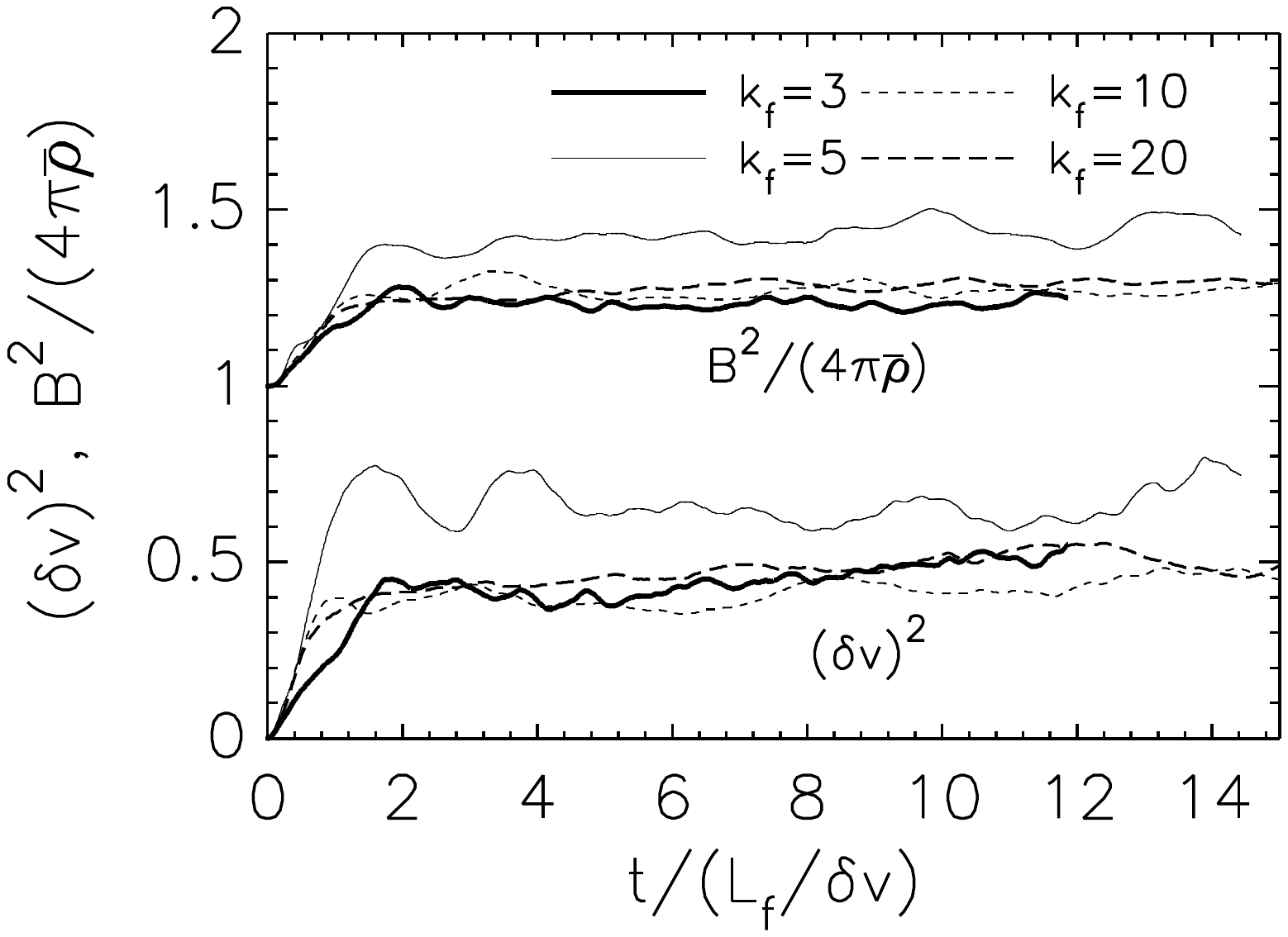}  
\hspace{3mm} 
\includegraphics[width=0.5\textwidth]{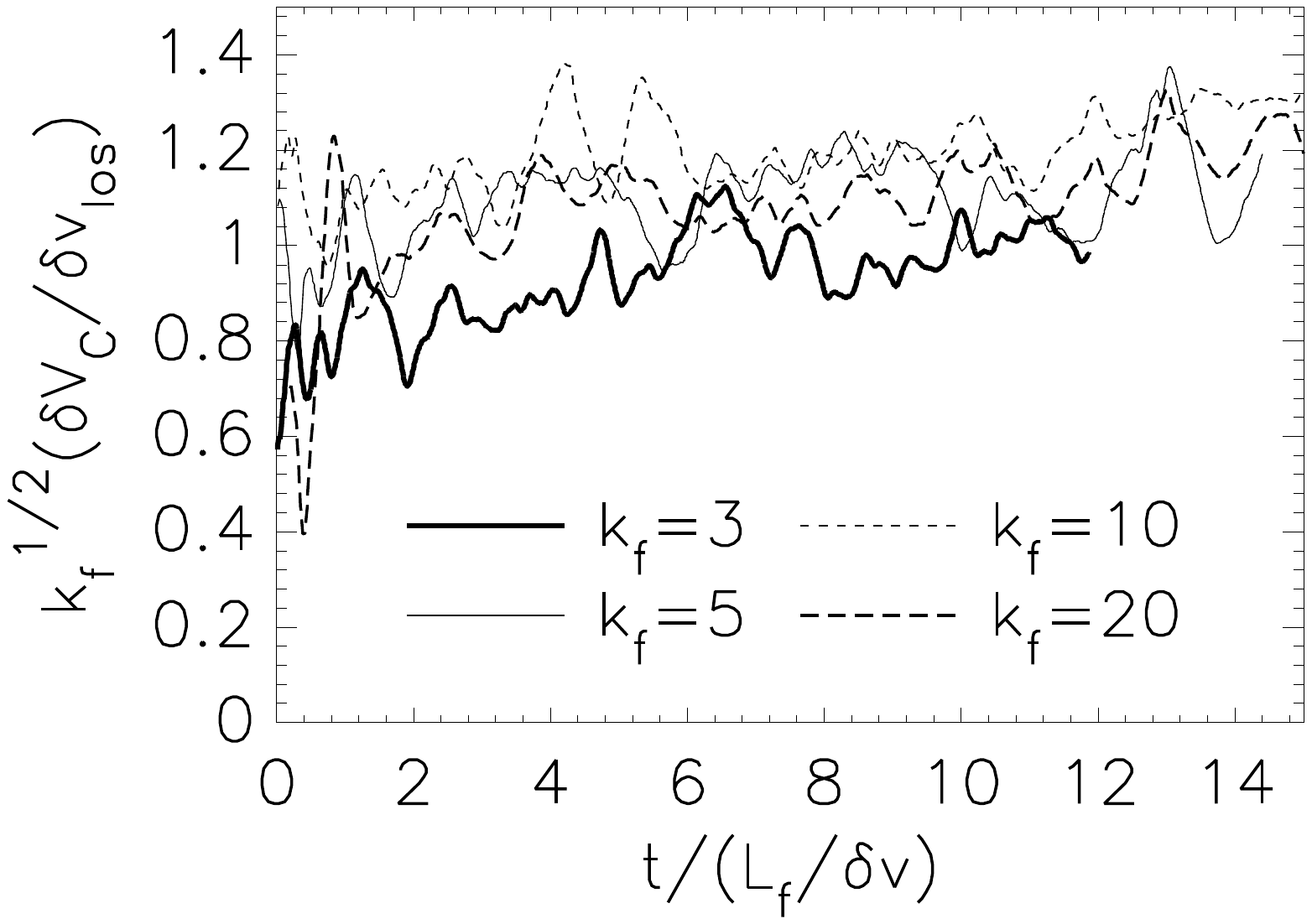}  
\caption{{\it Left:} Time evolution of $(\delta v)^2$ and $B^2/(4 \pi \bar{\rho})$.
  {\it Right:} Time evolution of 
     $\sqrt{k_f}$ times the standard deviation of velocity centroids $\delta V_c$
    divided by 
    the average line-of-sight velocity dispersion $\delta v_{los}$,  
   where $k_f$ is the driving wavenumber 
     (see Equation (\ref{eq:k_f}) for details).
   In both plots, the x-axes denote time normalized by $L_f/\delta v$, where $L_f$
   is the driving scale of turbulence and $\delta v$ is the r.m.s. velocity.
 }
\label{fig:3}
\end{figure*}

Indeed, our simulations confirm that, the larger the number of
independent eddies along the line of sight $L_{los}/L_f $ is, the smaller 
the variation  $\delta \phi$ is. 
Using our MHD turbulence data (see Table 1)
and numerical method in Fiege \& Pudritz (2000; see also Heitsch et al. 2001),
we obtain synthetic polarization maps arising from
magnetically-aligned dust grains at a far-infrared/sub-mm wavelength.
 We show the maps for runs with different driving scales  in Figure \ref{fig:2}.
 The coordinate system we adopted is similar to the one
 in Figure \ref{fig:1}: we assumed the mean magnetic field is along the
$x$-axis and the observer's line of sight is along the $z$-axis.
Since the mean magnetic field is along the $x$-axis, the direction of polarization is
mainly along the $y$-axis.    
 In the left panel, the driving scale is 
 $\sim$5 
                    times smaller than the 
 size of the computational box 
 (Run KF5) 
 and
 in the right panel the driving scale is $\sim$20 times smaller than the 
 size of the computational box (Run KF20).
 We can clearly see that the variation of polarization angle in the left panel 
 ($N$$\sim$5) 
 is
 larger than that in the right panel ($N$$\sim$20), which illustrates that
 the dispersion in polarization angle is indeed a function of $N$ ($\sim L_{los}/L_f$) as discussed
 in the previous paragraph.
 The standard deviations of polarization angle in 
 the left and right panels
 are 
 $\sim 9.1^\circ$    
 and
$\sim 3.5^\circ$, respectively.
 Note that the 3-dimensional magnetic fluctuations in both runs are similar 
 (see the left panel of Figure \ref{fig:3}).

\subsection{ Observational Estimation of ($L_f /L_{los}$) }  \label{sect:N}
From the discussion in the previous subsection,
it is clear that the conventional CF method overestimates the strength of
the mean plane-of-the-sky magnetic field by a factor of $\sqrt{L_{los}/L_f }$.
Therefore it is necessary to know $\sqrt{L_{los}/L_f }$ to obtain a correct
estimate of $B_{0,sky}$.

We propose that the standard deviation of centroid velocities $\delta V_c$ normalized by the average line-of-sight
velocity dispersion $\delta v_{los}$ is a good measure of $\sqrt{L_{los}/L_f}$ 
($\approx \sqrt{N}$):
\begin{equation}
  \frac{\delta V_c }{\delta v_{los}} \propto \frac{1}{\sqrt{N}},   \label{eq:sigC}
\end{equation}
where
\begin{equation}
    V_c =\int v_{los} I(v_{los}) dv_{los} \Bigg/ \int I(v_{los}) dv_{los}  \label{eq:vcent}
\end{equation}
and  $I(v_{los})$ is the observed line profile for the 
 line of sight.

Centroid velocity $V_c$ 
is a kind of average velocity. If we have several independent eddies
along the line of sight, then each eddy has its own mean velocity.
Then, roughly speaking, the observed centroid velocity for the line of sight is 
average of the mean velocities of individual eddies along the line of sight.
Note that the mean velocities of independent eddies are likely to be random.
Therefore, if we obtain centroid velocities for many different lines of sight and
calculate standard deviation of them, then
the standard deviation should be proportional to $\sim 1/\sqrt{N}$, where
$N$ is the number of independent eddies along a line of sight.
In Figure \ref{fig:1}(a) we draw 4 independent eddies along the line of sight
and in Figure \ref{fig:1}(c), we plot velocity profiles 
(or emission line profiles) from 4 independent eddies in solid lines and
the observed line profile in dashed line.
Then, the observed  centroid velocity $V_c$ is approximately the average of the
mean velocities of 4 eddies.
If we have more eddies along the line of sight, the variation in 
velocity centroids normalized by the average width of an optically thin emission line will become smaller.

\section{Results}
The left panel of Figure \ref{fig:3} shows the time evolution of 
$(\delta v)^2$ and $B^2/(4 \pi \bar{\rho})$, where
$B^2 = B_0^2+(\delta b)^2$.
At t=0, $(\delta v)^2 \approx 0$ and only the uniform magnetic field of
unit Alfv\'en speed (i.e.~$B_0/\sqrt{ 4 \pi \bar{\rho}}=1$) exists.
Due to driving, $(\delta v)^2$ and $B^2/(4 \pi \bar{\rho})$ grows initially.
After $t \sim 2(L_f/\delta v)$, turbulence  seems to show saturation in all simulations.

The right panel of Figure \ref{fig:3} shows the time evolution of
\begin{equation}
    \sqrt{k_f} \delta V_c / \delta v_{los} ~~(\approx \sqrt{N}  \delta V_c / \delta v_{los}  ), \label{eq:k_f}
\end{equation}
where 
$\delta v_{los}$ is
    the  average line-of-sight velocity dispersion.
The standard deviation of centroid velocities  $\delta V_c$ and 
the  average of line-of-sight velocity dispersion $\delta v_{los}$ are calculated over $512^2$
lines of sight perpendicular to the mean field $\bf{B}_0$.
The figure confirms that the quantity $\delta V_c / \delta v_{los}$ is indeed proportional
to $1/\sqrt{k_f}$ ($\approx \sqrt{L_f/L_{los}}$), as we proposed in Section \ref{sect:N}
(see Equation (\ref{eq:sigC})).

\begin{figure*}
\includegraphics[width=0.5\textwidth]{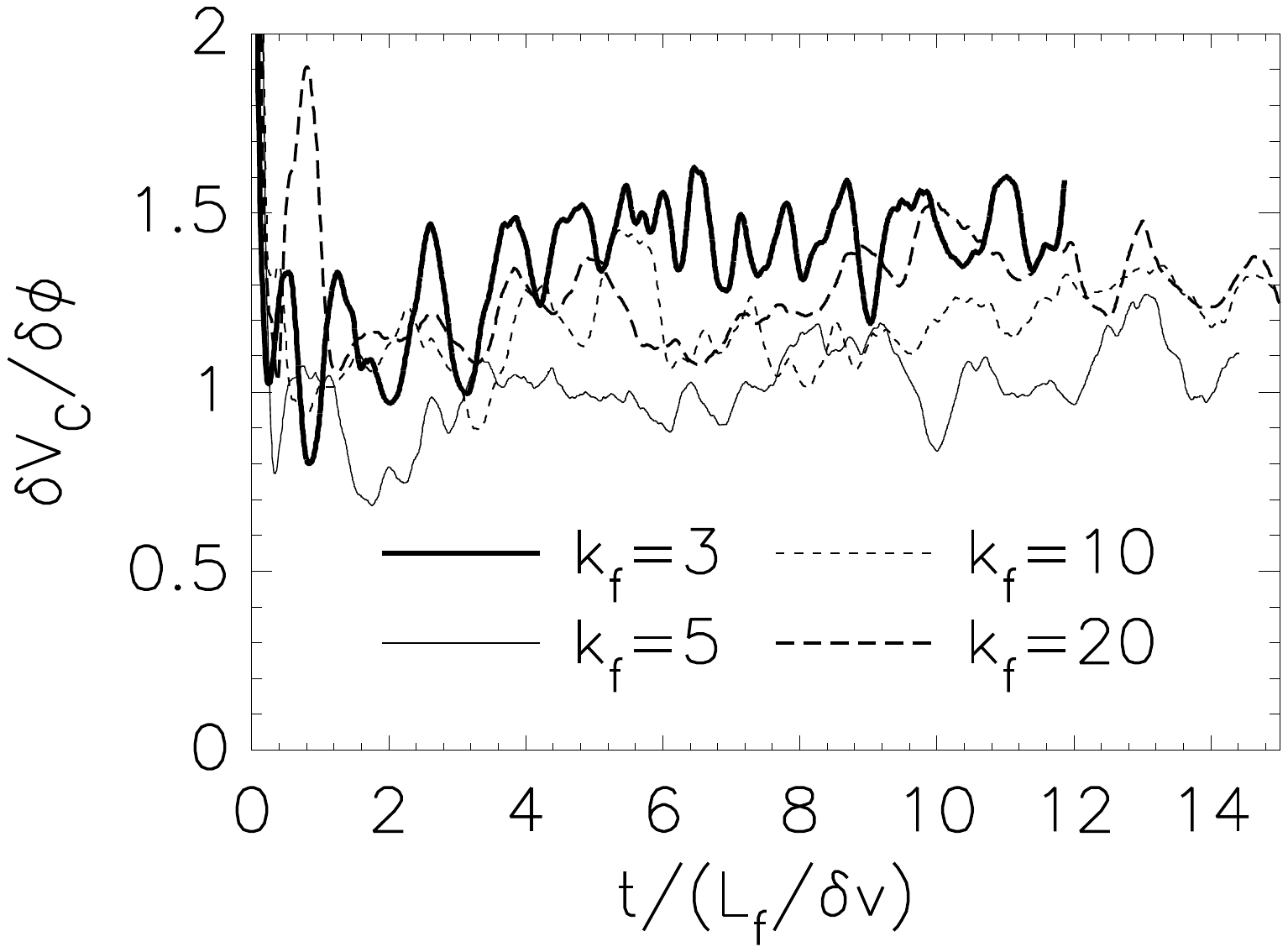}  
\hspace{3mm}
\includegraphics[width=0.5\textwidth]{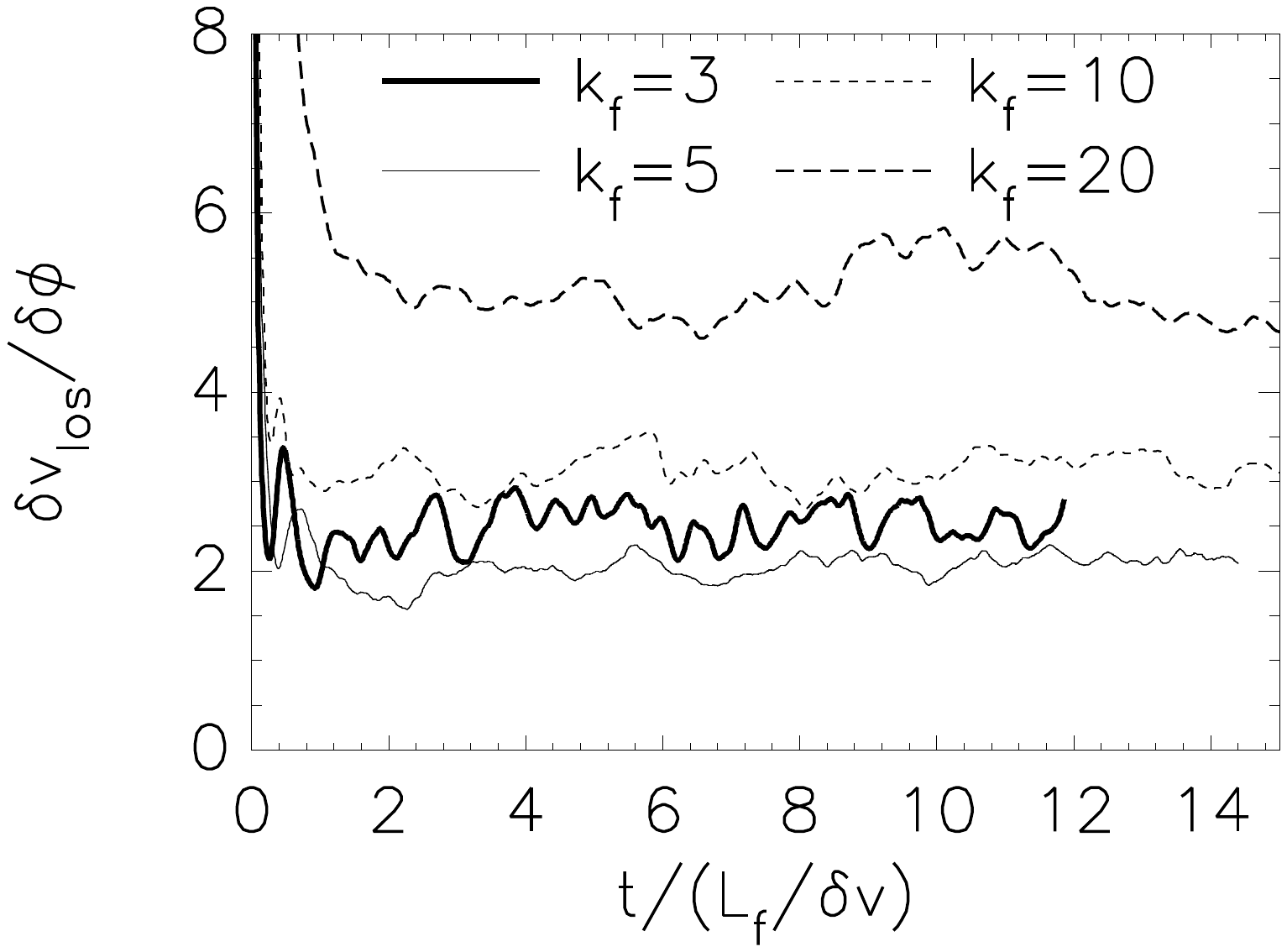}  
\caption{Estimates of $B_{0,sky}/\sqrt{4 \pi \bar{\rho}}$.
   {\it Left:} Results from our modified CF method (see Equation (\ref{eq:main2})).
   {\it Right:} Results from the conventional CF method (see Equation (\ref{eq:trad2})).
     Note that, in all our simulations, the line of sight is perpendicular to
     the mean field $\textbf{B}_0$ and $B_{0,sky}/\sqrt{4 \pi \bar{\rho}}$=1.
    The conventional CF method overestimates $B_{0,sky}$ when there are 
   many independent eddies along the line of sight.
 }
\label{fig:4}
\end{figure*}

The left panel of Figure \ref{fig:4} shows our main result.
Since the conventional CF method tends to overestimate $B_{0,sky}$ by a factor of 
$\sqrt{L_{los}/L_f}$ and $\delta V_c / \delta v_{los} \propto \sqrt{L_f/L_{los}}$ we can write
\begin{equation}
 B_{0,sky} = \xi^\prime \sqrt{ 4 \pi \bar{\rho} } \frac{ \delta V_c }{ \delta \phi } , \label{eq:main1}
\end{equation}
where $\xi^\prime$ is a constant of order unity that can be determined by numerical
simulations.
In the left panel of Figure \ref{fig:4} we plot estimates of 
$B_{0,sky}/\sqrt{4 \pi \bar{\rho}}$ from
this modified CF method:
\begin{equation}
    \frac{ \delta V_c}{ \delta \phi }.   \label{eq:main2}  
\end{equation}
In the panel, we can see that the estimates are fluctuating between $\sim$1.0 and $\sim$1.5.
Therefore, since $B_{0,sky}/\sqrt{4 \pi \bar{\rho}}=1$ in our simulations, the constant $\xi^\prime$ in Equation (\ref{eq:main1})
is between $\sim$0.7 and $\sim$1.

The right panel of Figure \ref{fig:4} shows estimates of 
$B_{0,sky}/\sqrt{4 \pi \bar{\rho}}$ from
the conventional CF method:
\begin{equation}
       \frac{ \delta v_{los} }{\delta \phi }.   \label{eq:trad2}
\end{equation}
We can see that the conventional CF method indeed overestimates $B_{0,sky}$
when the number of independent eddies along the line of sight ($\sim k_f$ in our simulations)
is large.
Note, however, that the conventional CF method seems to work fine for small $N$'s.

\section{Discussions and Summary}
In this paper, we have considered the effects of driving scale on the estimates of the mean
plane-of-the-sky magnetic field $B_{0,sky}$  from the Chandrasekhar-Fermi (CF) method.
The method we propose in Equation (\ref{eq:main1}) with    
$0.7 \lesssim \xi^{\prime} \lesssim 1.0$ does not require new observations. That is,
the method is readily applicable for present observational
data. Apart from numerical constants, the only difference between 
our method and the conventional CF method is that our method requires the standard deviation of velocity centroids $\delta V_c$, while the conventional method requires 
average width of the emission line profiles $\delta v_{los}$.
The standard deviation of velocity centroids $\delta V_c$ can be easily obtained 
from existing optically-thin emission line profiles.
If such emission line profiles ($I(v_{los})$'s) 
are available for $n_{obs}$ lines of sight, then we 
need the following two steps to obtain $\delta V_c$:    
\begin{enumerate}
\item We calculate the centroid velocity $V_c$ (see Equation (\ref{eq:vcent})) for each line of sight. Let $V_{c,i}$ be the centroid velocity for line of sight $i$: 
\begin{equation}
  V_{c,i} =\int v_{los} I_i(v_{los}) dv_{los} \Bigg/ \int I_i(v_{los}) dv_{los},
\end{equation}
    where $I_i(v_{los})$ is the optically-thin emission line profile for the line of sight.
\item We calculate $\delta V_c$ from the formula 
\begin{equation}
     \delta V_c^2 \equiv \frac{1}{n_{obs}} \sum_{i=1}^{n_{obs}} V_{c,i}^2 
                       -\left( \frac{1}{n_{obs}} \sum_{i=1}^{n_{obs}} V_{c,i} \right)^2. 
\end{equation}
\end{enumerate}

The CF method is useful for obtaining strengths of the 
plane-of-the-sky magnetic fields in molecular clouds.
Since observations suggest existence of supersonic motions and 
strong magnetic fields in molecular clouds,
we have considered  only supersonic ($M_s\sim 7$) and marginally  sub-Alfv\'enic 
($M_A   \lesssim 1$) MHD turbulence in this paper.
It is possible that the constant $\xi^{\prime}$ in Equation (\ref{eq:main1})
depends on the the sonic
Mach number $M_s$.
But, we expect that the dependence is weak 
because the sonic Mach number does not play an important role
in our discussion in Section \ref{sect:N}.
Nevertheless, more parameter study is needed to determine the dependence of 
$\xi^{\prime}$ on $M_s$.
Another limitation of our current work is that we have considered only the case that
the mean magnetic field 
is perpendicular to the line of sight, which means that
the inclination angle (with respect to the plane of the sky) of the mean magnetic field is zero.
In principle, our method, as well as the conventional CF method, should work for an
 arbitrary inclination
angle, unless the inclination angle is very close to  
90$^\circ$ (see discussions in Ostriker et al. 2001; Heitsch et al. 2001; Falceta-Gon{\c c}alves et al. 2008).
We will address these issues elsewhere.   

In this paper we have demonstrated that the conventional CF method indeed over-estimates
the mean plane-of-the-sky magnetic field $B_{0,sky}$ by a factor of $\sqrt{N}$, where
$N$ is the number of independent eddies along the line of sight.
We have found that the standard deviation of centroid velocities divided by the average
line-of-sight velocity dispersion ($\delta V_c /\delta v_{los}$) is proportional to $1/\sqrt{N}$
(Equation (\ref{eq:sigC}) and the right panel of Figure \ref{fig:3}).
Therefore 
 Equation (\ref{eq:main1})
 with $\xi^\prime = 0.7\sim 1$ provides a better estimate for $B_{0,sky}$.

\acknowledgements
This  work is supported by the National R \& D Program through 
the National Research Foundation of Korea, 
funded by the Ministry of Education (NRF-2013R1A1A2064475).
We thank  Woojin Kwon, Ellen Zweibel and Alex Lazarian for useful discussions.
We also thank the hospitality of the Astronomy Department at the University of Wisconsin - Madison during the final stage
of writing.

\begin{deluxetable}{lccccrrr}
\tabletypesize{\scriptsize}
\tablecaption{Simulations.}
\tablewidth{0pt}
\tablehead{
\colhead{Run} & \colhead{Resolution} 
   & \colhead{$M_s$ \tablenotemark{a}}
   & \colhead{$B_{0}/\sqrt{ 4 \pi \bar{\rho}}$ \tablenotemark{b}} 
   & \colhead{$k_f$ \tablenotemark{c}} 
   & \colhead{ $L_{los}/L_f$ \tablenotemark{d} }
}
\startdata
KF3        & $512^3$ & $\sim$7 & 1  & 3  & $\sim 3$ \\ 
KF5        & $512^3$ & $\sim$8 & 1  & 5  & $\sim 5$ \\ 
KF10        & $512^3$ & $\sim$7 & 1  & 10 & $\sim 10$ \\ 
KF20        & $512^3$ & $\sim$7 & 1  & 20  & $\sim 20$ 
\enddata
\tablenotetext{a}{Sonic Mach number.}
\tablenotetext{b}{Alfv\'en speed of mean magnetic field.}
\tablenotetext{c}{Driving wavenumber.}
\tablenotetext{d}{$L_{los}$ is the system size and $L_f$ is the driving scale. In our simulations, $L_{los}$=2$\pi$ and $L_f \sim 2\pi/k_f$. Therefore, we have 
$N= L_{los}/L_f\sim k_f$.}
\label{table_1}
\end{deluxetable}

\end{document}